# Orbital-selective metal-insulator transition and gap formation above $T_C$ in superconducting $Rb_{1-x}Fe_{2-y}Se_2$

Zhe Wang[1], M. Schmidt[1], J. Fischer[1], V. Tsurkan[1,2], M. Greger[3], D. Vollhardt[3], A. Loidl[1], J. Deisenhofer[1*]

[1]Experimental Physics V, Center for Electronic Correlations and Magnetism, University of Augsburg, 86159 Augsburg, Germany
[2]Institute of Applied Physics, Academy of Sciences of Moldova, 2028 Chisinau, Republich of Moldova
[3]Theoretical Physics III, Center for Electronic Correlations and Magnetism, University of Augsburg, 86159 Augsburg, Germany
*e-mail: joachim.deisenhofer@physik.uni-augsburg.de

Understanding the origin of high-temperature superconductivity in copper and iron based materials is one of the outstanding tasks of current research in condensed matter physics. Even the normal metallic state of these materials exhibits unusual properties. They have been interpreted as ordering phenomena of spin or charge carriers prior to the formation of the superconducting state[1-5]. Recently, an orbital-selective Mott transition has been suggested to occur in the normal-state phase of iron-selenide superconductors[6-8], where some of the five *d*-orbitals independently undergo metal-insulator-like transitions. Here we report on a hierarchy of temperatures $T_c < T_{gap} < T_{met}$ in superconducting $Rb_{1-x}Fe_{2-y}Se_2$ observed by THz spectroscopy. Above $T_{met}$ = 90 K the material reveals semiconducting characteristics. Below $T_{met}$ a coherent metallic THz response emerges. This metal-to-insulator type, orbital selective transition is clearly indicated by an isosbestic point in the temperature dependence of the optical conductivity and dielectric constant at THz-frequencies. At $T_{gap}$ = 61 K a gap opens in the THz regime and then the superconducting transition occurs at $T_c$ = 32 K. This sequence of temperatures seems to reflect a corresponding hierarchy of the electronic correlations in the different bands.

The selenide iron-based superconductors offer the opportunity to explore superconductivity all the way from binary FeSe with $T_c$ = 8 K (ref. 9) and the alkaline doped $A_{1-x}Fe_{2-y}Se_2$ ($A$ = K, Rb, and Cs) with $T_c$ = 32 K (refs 10-12) to epitaxial single-layer FeSe films with $T_c$ = 65 K (refs 13,14). Critical temperatures up to 46 K are obtained by intercalation[15-17] and 48 K by applying pressure[18,19].

Superconductivity in $A_{1-x}Fe_{2-y}Se_2$ was shown to occur only in single crystals, where a mesoscopic separation into a superconducting phase with $T_c \approx$ 32 K and an antiferromagnetic phase with a Néel temperature $T_N \approx$ 550 K has been established[10-12,20-23] (see Fig. 1c). The superconducting phase is electron-doped with about 0.15 electrons per Fe atom[22]. Angle-resolved photoemission spectroscopy (ARPES) studies determined the Fermi-surface topology with large electron-like Fermi sheets $\delta$ at the M-points and small electron pocket $\kappa$ at the Γ-point[24] (see Fig. 1d). Isotropic superconducting gaps of $2\Delta \simeq$ 16 meV and 20 meV were reported around the $\kappa$ and $\delta$ pockets, respectively[24]. In addition, a small gap $2\Delta \simeq$ 2 meV was observed by scanning tunneling spectroscopy[20]. Recently, the normal state of the superconducting phase was reported to exhibit an orbital-selective metal-to-insulator transition[7] observed by ARPES at about 100 K (ref. 6). Moreover, interphase correlations between superconducting and antiferromagnetic layers were predicted to reduce $T_c$ from a hypothetical 65 K to the actually observed value of 32 K due to proximity effects[25] (see Fig. 1e).

Figures 1a and 1b show the dielectric constant $\varepsilon_1$ and the optical conductivity $\sigma_1$ as a function of incident photon energy, respectively, derived from the time-domain THz transmission signal (see Methods) for selected temperatures across the three characteristic temperatures $T_c$ = 32 K, $T_{gap}$ = 61 K, and $T_{met}$ = 90 K. From room temperature down to $T_{met}$ = 90 K, $\varepsilon_1$ is positive and slightly increases towards lower temperatures, which is characteristic of a semiconducting optical response (Fig. 1a). Below 90 K, the dielectric constant becomes negative with a zero-crossing point from 2.5 meV at 80 K to 6.5 meV at 8 K. This is a fingerprint of a metallic system, and the observed zero-crossing points in $\varepsilon_1$ correspond to the reported screened plasma frequencies (refs. 21,26-28).

Similar observations, namely the smooth transition from an insulating to a metallic state, are made in the frequency dependence of the optical conductivity $\sigma_1$ for several temperatures (Fig. 1b). Above $T_{met} = 90$ K, $\sigma_1$ is low and almost energy independent, indicating a semiconducting behavior of quasi-localized charge carriers, while below this characteristic temperature a metallic response evolves with an increase of $\sigma_1$ towards lower energies. In contrast to the monotonous decrease of $\varepsilon_1$ with decreasing temperature, $\sigma_1$ decreases for temperatures below $T_{gap} = 61$ K and below energies of 5 meV. Towards lower temperature a gap-like feature (see arrow in Fig. 1b) evolves with a minimum in $\sigma_1$ close to $E_{gap} \sim 3$ meV. Its temperature dependence is shown as an inset to Fig. 2b. The gap energy appears to increase continuously across $T_c = 32$ K. Furthermore, the spectra of $\varepsilon_1$ and $\sigma_1$ do not exhibit drastic changes at $T_c = 32$ K, well below $T_{gap} = 61$ K.

To analyze these striking temperature induced changes of the optical response in more detail, Figs 2a and 2b show the dielectric constant and the conductivity as a function of temperature for several photon energies. At high temperatures, the dielectric constant is positive and almost temperature independent. At the same time the optical conductivity is low and increases gradually with decreasing temperature. This increase of $\sigma_1$ becomes stronger when $T_{met} = 90$ K is approached from above, the dielectric constant $\varepsilon_1$ decreases and becomes negative below $T_{met}$, signaling metallic behavior. This metal-to-insulator like transition is indicated by two sharp isosbestic points where all curves intersect and are therefore strictly frequency independent. At these points the dielectric constant and optical conductivity take the values $\varepsilon_1^{iso} = 14$ and $\sigma_1^{iso} = 22$ $\Omega^{-1}$cm$^{-1}$, respectively.

The increase of the conductivity for decreasing temperature terminates at $T_{gap} = 61$ K, and $\sigma_1$ starts to decrease, thus leading to a pronounced maximum. This signals the opening of a gap (see Fig. 1b) far above the superconducting transition temperature, $T_c = 32$ K. Indeed, the superconducting transition at $T_c = 32$ K leads to a weak kink (see inset of Fig. 2a for $\varepsilon_1$) in and a subsequent levelling off of both quantities.

The existence of isosbestic points in the plots of $\varepsilon_1(T, \omega)$ vs. $T$ and $\sigma_1(T, \omega)$ vs. $T$, respectively, allows one to determine their frequency dependences as $\varepsilon_1(T, \omega) = \varepsilon_1(T, \omega_0) + (1/\omega - 1/\omega_0) E_1(T)$ and $\sigma_1(T, \omega) = \sigma_1(T, \omega_0) + (\omega - \omega_0) S_1(T)$ in the vicinity of these points[29,30] (see Methods). Using the parameters $E_1(T)$ and $S_1(T)$ (Figs 3c and 3d), which were obtained by fitting the spectra as shown in Figs 3a and 3b, the validity of the approximations is demonstrated in Figs 3e and 3f. All curves are shown to collapse onto the one for $\omega_0 = 1.88$ meV. We note that the observed frequency dependencies can be applied to both temperature regimes, above and below $T_{met} = 90$ K. Given the intrinsic phase separation of the compound these frequency dependences are excellent low-frequency parametrizations of the actual optical response functions. Such well-defined frequency dependencies have not been observed in previous studies[26-28] and are valid only in the vicinity of sharp isosbestic points[29].

The change from insulating to metallic behaviour is reflected by the sign change of $E_1(T)$ and $S_1(T)$ from positive to negative at $T_{met} = 90$ K. While $E_1(T)$ saturates below $T_c$, $S_1(T)$ drops strongly with decreasing temperature until a minimum is reached at $T_{gap} = 61$ K. The gap-like suppression of the optical conductivity below $T_{gap}$ leads to a clear deviation from a linear frequency dependence, implying that the frequency dependence of $\sigma_1$ derived above does not hold below about 40 K.

The emergence of a metallic optical response below $T_{met} = 90$K can be understood as a consequence of an orbital-selective Mott transition, which was proposed to explain the result of a recent ARPES study in $A_x Fe_{2-y} Se_2$ with $A$ = K, Rb (ref. 6). Namely, for a doping of 0.15 electron per Fe, the investigation of a five-orbital Hubbard model discovered an orbital-selective Mott transition, where the $d_{xy}$ band contributes to the metallic properties only below a temperature of about 100 K, while the $d_{xz}/d_{yz}$ bands retain their metallic features both below and above 100 K (refs 6,7). This scenario closely follows the general physical picture of strong orbital differentiation with a low coherence temperature[31].

The collapsed optical conductivity curve in Fig. 3f exhibits the same temperature dependence as the ARPES spectral weight associated with the $d_{xy}$ band of the electron pocket at the M-point[6]. Thus we conclude that the optical conductivity probes predominantly the $d_{xy}$ band and signals a Mott transition of this band at the temperature of the isosbestic point $T_{met} = 90K$.

The pronounced orbital differentiation (ref. 8) of $Rb_{1-x}Fe_{2-y}Se_2$ manifests itself through highly orbital-dependent mass renormalizations of about 10 for the $d_{xy}$ band and about 3 for the $d_{xz}/d_{yz}$ bands, respectively[6]. Owing to the significantly stronger correlation strength, the lifetime of quasiparticles in the $d_{xy}$ band is much more susceptible to temperature than in the other bands. This leads to a selective reduction of the $d_{xy}$ quasiparticle peak, which effectively eliminates its contribution to transport processes with increasing temperature. The lighter $d_{xz}/d_{yz}$ quasiparticles will, however, dominate the transport properties. Indeed, no anomaly is visible in the $dc$-resistivity at $T_{met} \approx 90$ K (ref. 12). The optical conductivity is expected to be governed by all quasiparticles, but instead it determines the temperature dependence of the strongly renormalized $d_{xy}$ charge carriers.

To explain the orbital differentiation of the optical conductivity, we have to consider the special morphology of the phase-separated system. Huang and coworkers investigated effects of interlayer hopping at the interface of the superconducting and the antiferromagnetic phases in the $d_{xz}/d_{yz}$ channels (sketched in Fig. 1e) and found that the resulting distortions of the Fermi surfaces effectively reduce $T_c$ (ref. 25). We assume that these incoherent hopping processes via the $d_{xz}/d_{yz}$ bands lead to large scattering rates in the optical response of the corresponding quasi-particles of the $d_{xz}/d_{yz}$ bands. By contrast, the $d_{xy}$ channel remains almost unaffected by the proximity effect and reveals its metallic optical response at low-frequencies via its larger mass normalization.

Without the distortions of the Fermi surface induced by the vicinity of the antiferromagnetic phase a value of $T_c$ as high as 65 K is estimated[25]. This scenario is supported by the formation of a gap already below $T_{gap} = 61$ K in the $d_{xy}$-dominated optical conductivity. The upper bound for the gap at 3.2 meV must be compared to the gap value of $2\Delta \simeq 2$ meV found in a scanning tunneling microscopy study on the $K_{1-x}Fe_{2-y}Se_2$ films[20]. The fact that the conductivity is not completely suppressed below 3.2 meV points towards an anisotropic nature of this low-frequency gap.

Our observations provide strong evidence for the existence of an orbital-selective Mott transition, and also point to the possibility of orbital-selective superconducting properties[32]. The appearance of a gap in the $d_{xy}$-dominated optical conductivity at $T_{gap} = 61$ K in superconducting $Rb_{1-x}Fe_{2-y}Se_2$ may be compared with the reported opening of a gap below 65 K in single-layer FeSe films, which was interpreted as an indication for a superconducting transition at this temperature[13]. Given the similarity of the electronic band structures of FeSe mono-layer films and $Rb_{1-x}Fe_{2-y}Se_2$ (ref. 13), we predict that the highest possible value of $T_c$ in Fe selenide systems is determined by the correlated quasiparticles in the $d_{xy}$ channel.

**Methods**

**Crystal growth.** 99.75% pure Rb and polycrystalline FeSe, preliminarily synthesized from the high-purity elements (99.75% Fe and 99.999% Se) were used as starting materials. Crystals of $Rb_{0.74}Fe_{1.60}Se_2$ were grown using Bridgman method from the starting composition corresponding to the $Rb_{0.8}Fe_2Se_2$ stoichiometry. The samples have soaked at 1070 °C for 5 hours. The cooling rate is 3 mm per hour (ref. 12).
Crystals of the same sample batch were investigated by magnetization (ref. 12), resistivity, nuclear magnetic resonance (ref. 22), muon spin rotation and scanning-near field microscopy (ref. 23).

**Scaling analysis around the isosbestic points.** In general, an isosbestic point is an intersection point of a family of $n$ curves $f(x, p_i)$, $i = 1, 2, ..., n$ in the plot of $f(x)$ (refs 29,30). Since isosbestic behaviour is observed only in a certain parameter range around a particular value $p_0$, we can expand around $p_0$ as $f(x, p_i) = f(x, p_0) + (p_i - p_0) F_1(x, p_0) + O[(p_i - p_0)^2]$, where $F_1(x) = \partial f/\partial p|_{p=p_0}$ is a function of temperature only. The scaling of $f(x, p_i)$ is calculated as $\tilde{f}(x, p_i) = f(x, p_i) - (p_i - p_0) F_1(x, p_0) = f(x, p_0) + O[(p_i - p_0)^2]$. The validity of the expansion approximation is verified by $\tilde{f}(x, p_i)$ curves for different $p_i$ collapsing on a single curve. Here, $x \equiv T$; $f \equiv \varepsilon_1$, and $\sigma_1$; $\tilde{f} \equiv \tilde{\varepsilon}_1$, and $\tilde{\sigma}_1$; $F_1 \equiv E_1$, and $S_1$; $p_i \equiv 1/\omega_i$, and $\omega_i$, respectively.

**THz spectroscopy measurements.** Time-domain THz transmission measurements were carried out on a single crystal with the THz electric field parallel to the *ab*-plane in the spectral range 1 – 10 meV using a TPS spectra 3000 spectrometer (TeraView, Ltd.). Transmission and phase shift were obtained from the Fourier transformation of the time-domain signal. The dielectric constant and optical conductivity were calculated from the transmission and phase shift by modelling the sample as a single-phase dielectric slab. The single crystal for optical measurements was prepared with the thickness of about 45 μm and a cross section of about 5 mm$^2$. A $^4$He-flow magneto-optical cryostat (Oxford Instruments) was used to reach the temperature range 8 – 300 K.

**Acknowledgements**

We thank Dirk van der Marel, Alessandro Toschi, Christian Bernhard, and Alexander Boris for helpful discussions. This work was partially supported by the Deutsche Forschungsgemeinschaft via the Transregional Collaborative Research Centers TRR 80, Priority Program SPP 1458, and Project DE 1762/2-1.

**Author contributions**

J.D., A.L. and D.V. conceived and supervised the project. V.T. prepared the high-quality single crystals. J.F., M.S. and Z.W. performed the optical experiments. J.D., J.F., M.G. and Z.W. analysed the data. All authors contributed to the interpretation of the data and to the writing of the manuscript.

**Competing financial interests**

The authors declare no competing financial interests.

Figure 1

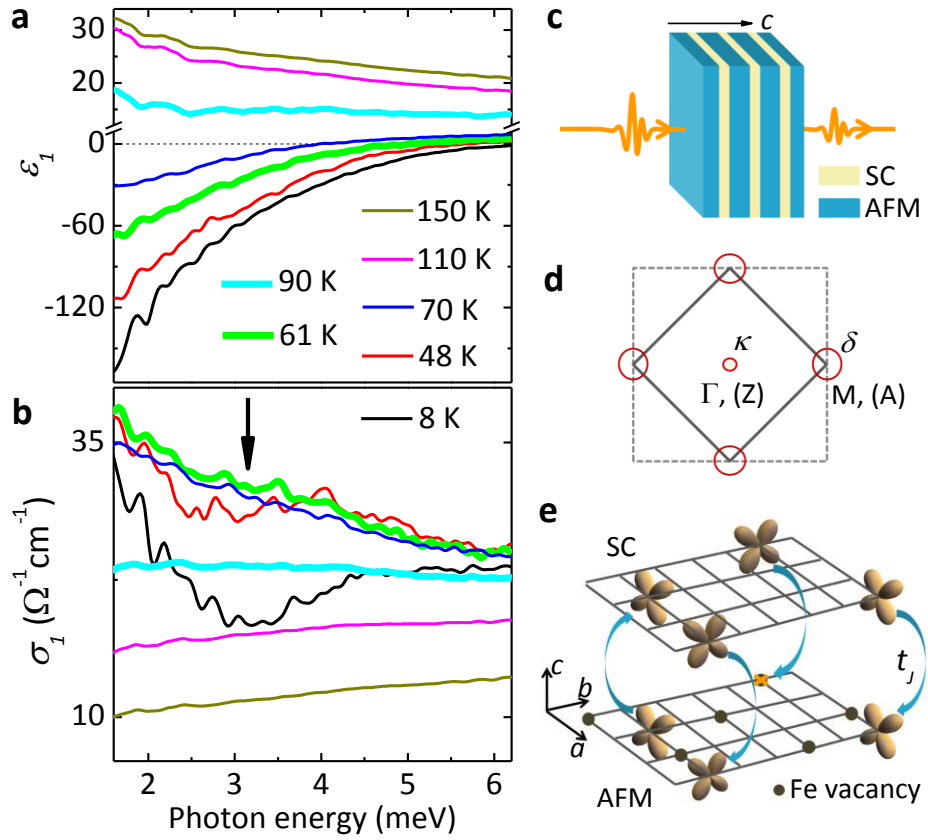

**Figure 1 | Photon energy dependence of dielectric constant and optical conductivity. a,b**, Dielectric constant $\varepsilon_1$ and optical conductivity $\sigma_1$ of $Rb_{0.74}Fe_{1.60}Se_2$ in the THz spectral region at various temperatures. The arrow in **b** indicates the formation of a gap at about 3 meV below $T_{gap} \approx 61$ K. **c**, Schematic sketch of the THz transmission experiment on the sample consisting of superconducting (SC) and antiferromagnetic (AFM) layers[23]. **d**, Sketch of Brillouin zone and Fermi surface[24]. **e,** Illustration of interphase $d_{xz}/d_{yz}$ electron hopping ($t_J$) between Fe atoms from SC and from AFM layers (adapted from Ref. 25).

Figure 2

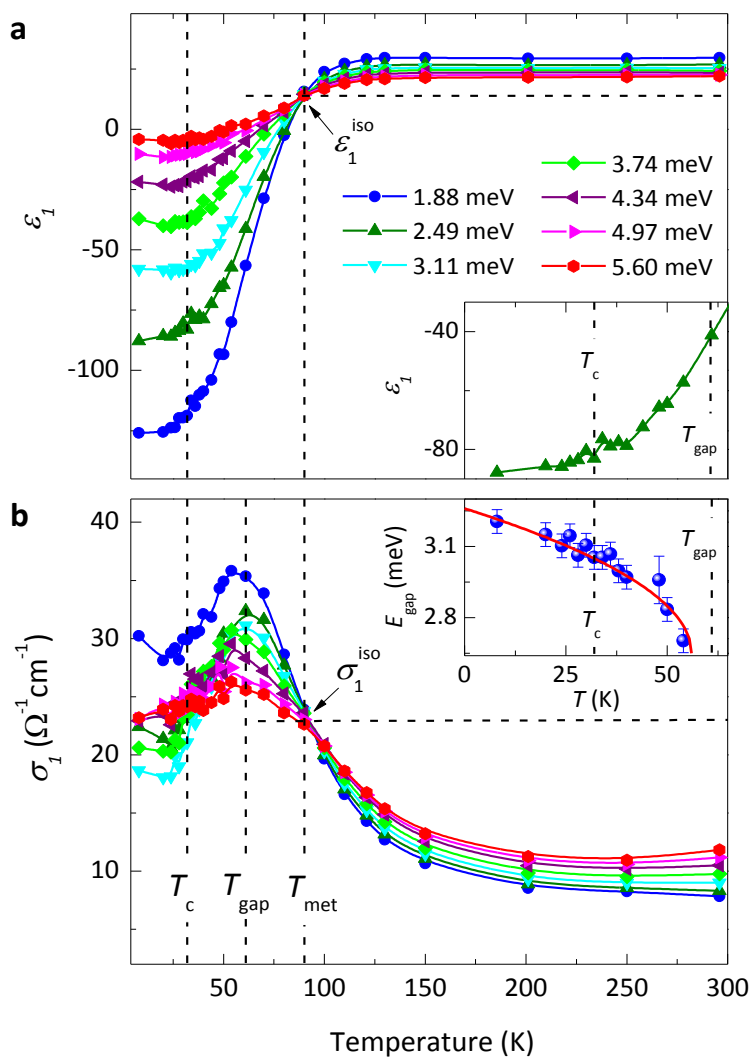

**Figure 2 | Temperature dependence of dielectric constant and optical conductivity. a,b,** Dielectric constant $\varepsilon_1$ and optical conductivity $\sigma_1$ as a function of temperature at different photon energies. The characteristic temperatures and the isosbestic points are indicated by dashed lines. Inset to **a**, temperature dependence of $\varepsilon_1$ at 2.49 meV around $T_c$. Inset to **b**, temperature dependence of the gap energy $E_{gap}$ (the solid line is drawn to guide the eyes).

Figure 3

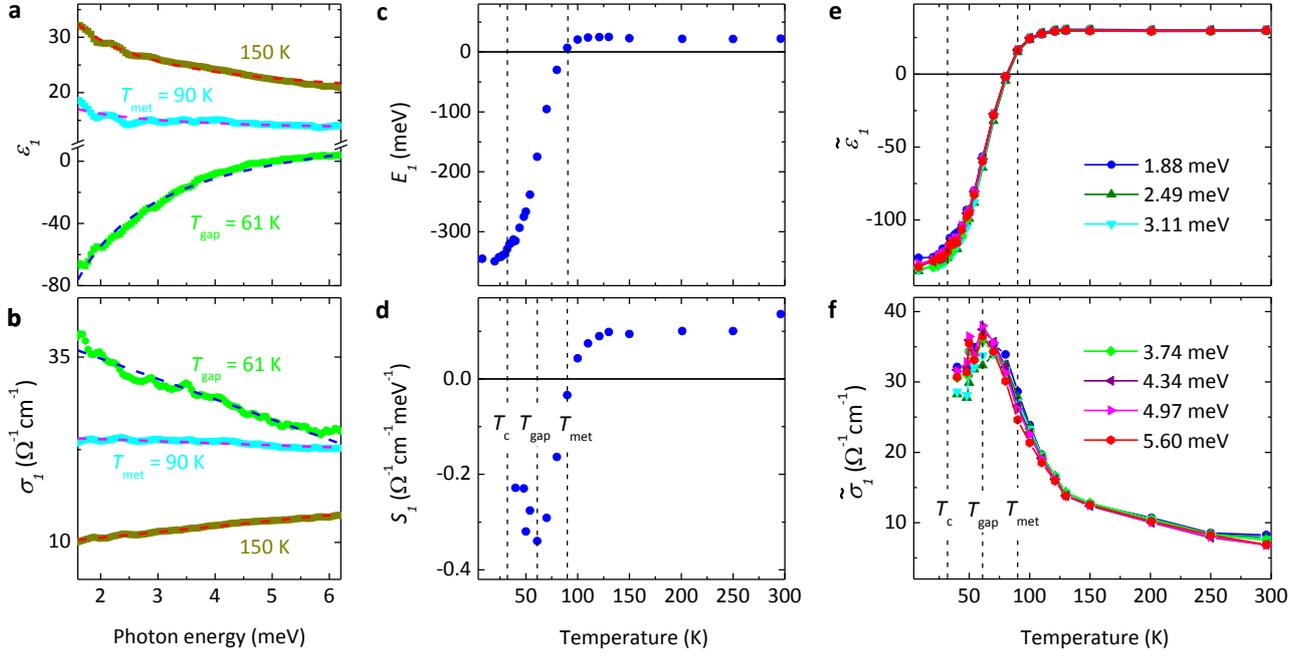

**Figure 3 | Scaling of the dielectric constant and the optical conductivity. a,b,** Fits of the dielectric constant using $\varepsilon_1(\omega) \sim E_1/\omega$ and $\sigma_1(\omega) \sim S_1\omega$ for selected temperatures, respectiviely. **c,d,** Temperature dependences of the respective fit parameters $E_1$ and $S_1$. **e,f,** Scaled dielectric constant $\tilde{\varepsilon}_1 = \varepsilon_1 - (1/\omega_i - 1/\omega_0) E_1$ and optical conductivity $\tilde{\sigma}_1 = \sigma_1 - (\omega_i - \omega_0) S_1$ for different frequencies $\omega_i$. The curves collapse on the ones for $\omega_0$ = 1.88 meV (see Methods).

13